\documentclass[twocolumn,trackchanges]{aastex701}

\submitjournal{ApJ}

\shorttitle{}
\shortauthors{Patel et al.}

\defcitealias{Bennet_2024}{B24}
\defcitealias{bennet_patel_2025}{BP25}
\begin{document}

\title{Encounters Between M33 and Present-Day M31 Satellites Hint at a Previous Group Accretion}

\author[0000-0002-9820-1219]{Ekta~Patel}\affiliation{Department of Astrophysics and Planetary Sciences, Villanova University, 800 E. Lancaster Ave, Villanova, PA 19085, USA}
\email[show]{ekta.patel@utah.edu}
\correspondingauthor{Ekta Patel}

\author[0000-0001-8354-7279]{Paul Bennet}
\affiliation{Space Telescope Science Institute, 3700 San Martin Drive, Baltimore, MD 21218, USA}
\email{pbennet@stsci.edu}

\author[0000-0001-8368-0221]{Sangmo Tony Sohn}
\affiliation{Space Telescope Science Institute, 3700 San Martin Drive, Baltimore, MD 21218, USA}
\affiliation{Department of Astronomy \& Space Science, Kyung Hee University, 1732 Deogyeong-daero, Yongin-si, Gyeonggi-do 17104, Republic of Korea}
\email{tsohn@stsci.edu}

\author[0000-0003-4207-3788]{Mark A.\ Fardal}
\affiliation{Eureka Scientific, 2452 Delmer Street, Suite 100, Oakland, CA 94602, USA}
\email{mfardal@eurekasci.com}

\author[0000-0001-7827-7825]{Roeland P. van der Marel}
\affiliation{Space Telescope Science Institute, 3700 San Martin Drive, Baltimore, MD 21218, USA}
\affiliation{Center for Astrophysical Sciences, The William H. Miller III Department of Physics \& Astronomy, Johns Hopkins University, Baltimore, MD 21218, USA}
\email{marel@stsci.edu}

\begin{abstract}

This work investigates whether two known Andromeda (M31) satellites, Pisces (LGS~3) and Andromeda XVI, have interacted with M33, M31's most massive satellite. $\Lambda$CDM predictions imply a handful of satellite galaxies around M33, yet few M33 satellites have been found and confirmed despite its high mass. We use proper motions combined with backward orbit integration in a semi-analytic potential to constrain plausible interaction scenarios for Pisces and And~XVI. Both dwarfs are currently M31 satellites, defined as being inside its virial radius. However, our results show that, in our fiducial mass models, 42\% (And~XVI) and 60\% (Pisces) of dwarf orbits support that they were previously satellites of M33 (i.e., once inside its virial radius). Both dwarfs had fly-by encounters with M33 at relative velocities greater than M33's escape speed within the past 1-2 Gyr. In over 70\% of orbits, Pisces and And~XVI also had a close approach with each other post-M33 interaction and share an orbital plane, suggesting possible past group accretion. We explore a range of mass combinations for M31 and M33, finding that this primarily regulates the likelihood that the dwarfs were satellites of M33 in the past, while upholding conclusions of recent flybys about M33. These close interactions provide new evidence for past satellite exchange and/or group infall scenarios between M31 and M33. Such interactions also affect comparisons to observational surveys that define satellites primarily by their distance relative to host galaxies.

\end{abstract}

\keywords{Dwarf Galaxies (416), Local Group (929), Galaxy Dynamics (591)}

\section{Introduction} \label{sec:intro}

In the $\Lambda$ Cold Dark Matter ($\Lambda$CDM) paradigm, galaxies assemble hierarchically. As low mass galaxies accrete onto the halos of more massive galaxies, many of the low mass galaxies become satellite galaxies of their massive companion \citep[][]{Moore_1999,Gao_2004,Kravtsov_2004,Bullock_2017}. Satellite galaxies are predicted to be self-similar in $\Lambda$CDM, such that massive dwarf galaxies also host a population of satellite galaxies. However, historically, satellites of dwarfs have been hard to find observationally \citep[][]{Munshi2019}. 

Nonetheless, within our Galaxy, there has been recent progress in the number of ultra-faint dwarfs (UFDs), discovered around the LMC \citep[e.g.,][]{bechtol15,dwagner15,koposov15}. Subsequent studies have confirmed a dynamical association between these UFDs and the LMC as satellites that entered the halo of the Milky Way (MW) along with the accretion of the LMC \citep[e.g.,][]{Erkal_2020, Patel_2020}. 

Just beyond the Local Group (LG), resolved, low mass dwarf satellites have also been detected around NGC~3109 \citep[][]{Hargis_2020,Doliva-Dolinsky_2025} and NGC~2403 \citep[][]{Carlin_2024}, two LMC-mass galaxies. Furthermore, a satellite in the process of active disruption around dwarf starburst galaxy NGC~4449 has also been observed \citep{Martinez_2012}, showcasing the hierarchical nature of galaxies at the dwarf-mass scale. Several more distant surveys are also expanding observational searches to unresolved dwarfs around massive dwarf hosts in the Local Volume \citep[][]{Davis_2024, Hunter_2025, Li_2025}. 

M33 is a notable exception to this recent success. The third largest galaxy in the LG has a very small or nonexistent satellite system \citep[][]{Patel_2018b}, despite being more massive than the LMC, and many of the Local Volume systems successfully searched for dwarf satellites \citep[][]{McConnachie_2012}. Furthermore, recent kinematic studies have shown that M33 is likely on first infall into M31 \citep[][]{Patel_2017a, vdmarel2019, Patel_2025, Wu_2025} and therefore surviving satellites should remain bound to M33 at the present day, contrary to previous work suggesting a recent tidal interaction between M33 and M31 during which M33 satellites may have been stripped \citep{McConnachie_2009, Putman_2009}. Despite this compelling evidence that M33 too should host a population of low mass dwarfs \citep{Patel_2018b}, like the LMC, deep surveys of M33's halo have only explored about 35\% of M33's virial volume\citep[assuming a virial radius of 161 kpc;][]{McConnachie_2009, Patel_2018b}.

To date, only three candidate M33 satellite galaxies have been reported: UFD And~XXII \citep{Martin_2009, Chapman_2013}, UFD Pisces~VII/Triangulum~III \citep{Martinez_2022, Collins_2024}, and ultra-diffuse dwarf Triangulum~IV \citep{Ogami_2024}. Without available distances, proper motions (PMs), and line-of-sight (LOS) velocities ($\rm V_{LOS}$) for each of these dwarfs, it is not yet possible to determine whether these galaxies are dynamically associated with M33. 

In addition, only And~XXII has a reported star formation history \citep{Savino_2025}, which shows that it quenched about 4.5 Gyr ago; however, it formed a majority ($\sim$80\%) of its stars by about 10 Gyr ago. \citet{Collins_2024} shows tentative evidence for recent star formation in Pisces VII between 1-2 Gyr ago. This recent star formation could be associated with a possible past interaction with M33; however, as noted in \citet{Collins_2024}, the possibility that the young stars are blue stragglers remains. Interactions with another galaxy may explain the dynamically heated nature of Tri IV. While the potential perturber of Tri~IV is yet unknown, \citet{Ogami_2024} has suggested M33 is a strong candidate.

Here we report on new orbital results for Pisces (LGS3) and Andromeda~XVI (And~XVI), which are presently considered M31 satellite galaxies, but may have previously interacted with M33 based on reconstructed orbital histories. If so, these dwarfs would fill in a key part of the M33 satellite mass function, providing a classical dwarf (Pisces) and a dwarf on the boundary between classical and ultra-faint (And~XVI). This possible transfer of satellite galaxies back and forth between M33 and M31, and massive satellites and their hosts more generally, has yet to be explored in detail. It may be indicative of remnants of group infall scenarios.

We focus on these particular dwarfs because orbital results from a companion paper \citep[][hereafter BP25]{bennet_patel_2025} revealed a compelling possibility of a past interaction between M33 and Pisces, based on backward orbit integrations that explicitly include M33's gravitational influence. Preliminary analysis of And XVI in the same orbital framework suggests it too experienced a close passage with M33. A more detailed discussion of And XVI's orbital history within the broader M31 satellite population will be presented in forthcoming work (Sohn et al., in prep.).

This paper is organized as follows. In Section \ref{sec:methods}, we briefly discuss the data and methods used to investigate the dynamics of Pisces and And~XVI. In Section \ref{sec:orbits}, we report orbital parameters for Pisces and And~XVI and explore possible evolutionary scenarios. Section \ref{sec:discussion} includes a discussion of the plausibility of group infall versus independent accretion events, the impact on M33's satellite mass function, and M33's morphology. Finally, we conclude in Section \ref{sec:conclusions}.

\section{Data and Methods}
\label{sec:methods}

\subsection{Galactocentric Coordinates}
\label{subsec:coords}
\begin{deluxetable*}{lccccccccc}
%\tablenum{1}
\tablecaption{Position and Velocity Vectors \label{tab:Pos_vel}}
\tablewidth{0pt}
\tablehead{
\colhead{Dwarf}  &  \colhead{$x$} & \colhead{$y$} & \colhead{$z$} &  \colhead{$v_x$} & \colhead{$v_y$} & \colhead{$v_z$} & \colhead{$v_{rad}$} &\colhead{$v_{tan}$} & \colhead{$v_{tot}$}\\
\multicolumn1c{} & \colhead{(kpc)} &  \colhead{(kpc)} & \colhead{(kpc)} &  \multicolumn1c{(km $s^{-1}$)} & \multicolumn1c{(km $s^{-1}$)}  & \multicolumn1c{(km $s^{-1}$)} & \multicolumn1c{(km $s^{-1}$)}  & \multicolumn1c{(km $s^{-1}$)}  & \multicolumn1c{(km $s^{-1}$)} 
}
%\decimalcolnumbers
\startdata
Andromeda~XVI & -264$\pm$9 & 366$\pm$13 & -262$\pm$10 & 108$\pm$85 & -36 $\pm$74 & 247$\pm$99 & -204$\pm$5 & 180$\pm$98 & 272$\pm$71 \\
Pisces & -282$\pm$6 & 366$\pm$8 & -395$\pm$9 & 84$\pm$26 & 43$\pm$23 & 210$\pm$22 & -149$\pm$3 & 175$\pm$29 & 230$\pm$22 \\
\enddata
\tablecomments{Three-dimensional position and velocity vectors derived, in Galactocentric coordinates, from the galaxy positions, distances, LOS velocities, and PM measurements as reported in \citetalias{bennet_patel_2025} for Pisces and Sohn et al., in prep. for And~XVI. The Galactocentric distance of the Sun and the circular velocity of the local
standard of rest (LSR) are adopted from \citet{mcmillan11}. Solar peculiar velocities are taken from \citet{Schonrich_2010}. }
\label{tab:6d}
\end{deluxetable*}

PMs are measured from two or more epochs of imaging separated by $N$ years, where the distance to the object of interest determines $N$. For And~XVI, the first and second epochs of imaging came from Hubble Space Telescope (HST) programs GO-13028 (2013, P.I.~E.~Skillman) and GO-16273 (2021, P.I.~S.~T.~Sohn), respectively. While for Pisces three epochs of imaging were obtained, the first as part of GO-10505 (2005, P.I.~C.~Gallart), the second as part of GO-13738 (2015, P.I.~E.~J.~Shaya), and the final as part of GO-17174 (2023, P.I.~P.~Bennet). 

PMs are derived using the same techniques as those described extensively in previous works \citep[][]{Sohn_2012, Sohn_2013, Sohn_2017, Sohn_2020, Bennet_2024}. Please see Sohn et al., in prep. for And~XVI and \citepalias{bennet_patel_2025} for Pisces, for all relevant data and reduction properties. 

As in \citetalias{bennet_patel_2025}, distances to M33 and M31 are adopted from \citet{Savino_2022}. PMs and LOS velocity for M33 are from \cite{Brunthaler_2005} and \citet{Corbelli_1997}, respectively. For M31, PMs are adopted from \citet{Salomon_2021} and LOS velocity is from \citet[][see also \citet{Slipher_1913}]{ vandenBergh99}.

For Pisces, we use $V_{LOS}=-287\pm1$ km s$^{-1}$ \citep{Huchtmeier_2003} and $D_{M31}=293\pm6$ kpc \citep{Savino_2022}. For And~XVI, $V_{LOS}=-367\pm3$ km s$^{-1}$ \citep{Tollerud_2012} and $D_{M31}=517.6\pm19$ kpc \citep{Savino_2022}. At face value, Pisces and And~XVI have $V_{LOS}$ values much closer to that of M31 ($V_{LOS}=-301$ km s$^{-1}$) than M33 ($V_{LOS}=-180$ km s$^{-1}$), and this has been used as evidence of their association to M31. However, when PMs are combined with literature distances and line-of-sight velocities ($V_{LOS}$) to derive full 6D phase-space information, as provided in Table \ref{tab:6d}, it becomes clear that the tangential velocities of both dwarfs are significant, warranting a closer investigation of the dynamics between these dwarfs and both M31 and M33. 

Uncertainties on each position and velocity component in Table \ref{tab:6d} represent the standard deviation of 1,000 samples drawn in a Monte Carlo fashion from the joint uncertainties on PMs, distances, and LOS velocities.

\subsection{Orbital Models}
\label{subsec:orbit_models}
To investigate the orbital dynamics of Pisces and And~XVI, we use the same modeling techniques as in \citetalias{bennet_patel_2025} and previous work \citep{Bennet_2024,Patel_2025, Patel_2020, Patel_2017a}. This orbital framework follows the center of mass evolution of the MW, M31, LMC, M33, and the dwarf of interest. All galaxies are represented as extended mass distributions with rigid potentials.

In brief, the MW and M31 are modeled as three-component potentials with a halo, disk, and bulge. We include dynamical friction owing to the MW and M31 and adiabatically contract the halos of the MW and M31 using the \texttt{CONTRA} code \citep{contra}. The following masses are adopted: $(2, 1, 0.25, 0.18) \times 10^{12}\,M_{\odot}$ for M31 \citep{patel23}, the MW \citep{wang20,mcmillan11}, M33 \citep{guo_10}, and the LMC \citep{watkins_24}, respectively.

Beginning with the Galactocentric coordinates in Table \ref{tab:6d}, equations of motions are integrated backward in time for a period of 6 Gyr with a Leapfrog algorithm. The integration follows the evolution of positions and velocities of the MW, M31, the LMC, M33, and each dwarf of interest. All galaxies exert gravitational forces on each other, with the exception of the dwarf, which feels the influence of other massive bodies but does not exert any force on them. Importantly, these five-body encounters account for the center of mass shifts in the position and velocity of the MW/M31 as a result of the passage of massive satellite galaxies, like the LMC/M33 \citep{gomez15, Patel_2025}.

%As noted in previous work, this analysis does not account for alternative masses for the MW and M31, which are known to cause significant degeneracies in orbital history derivations, equivalent to the magnitude of the uncertainties on PMs themselves \citep{dsouza22}. Future work will address how the assumed masses of the MW and M31 affect orbital uncertainties for Pisces and And~XVI.

The dwarfs are modeled as point masses with total halo masses determined from the \citet{moster13} abundance matching relation. And~XVI's stellar mass is $M_*=8 \times 10^4\, M_{\odot}$ \citep{McConnachie_2018}, which corresponds to a halo mass of approximately $M_{halo}=1 \times 10^9\, M_{\odot}$\footnote{It should be noted that there is significant scatter on the assumed halo masses for dwarf galaxies. While we do not account for these uncertainties here, the effect of different dwarf masses, while expected to be minimal, should be rigorously quantified in future work.}. Pisces' stellar mass is $M_*=9.6 \times 10^5\, M_{\odot}$ \citep{McConnachie_2012}, which corresponds to a halo mass of approximately $M_{halo}=5 \times 10^9\, M_{\odot}$.

We will refer to the orbital histories resulting from backward orbit integration using the average 6D phase space information (listed in Table \ref{tab:6d}) as direct orbital histories \citep[see][]{Patel_2020}. Uncertainties on orbital histories will be further discussed in Section \ref{sec:orbits}. We refer to the specific combination of masses mentioned above, specifically the values of adopted masses for M33 and M31, as our \emph{fiducial} scenario. In Section \ref{subsec:mass_prop_results}, we will discuss the implications of assuming different M33 and M31 masses.

In this work, we will use the same terminology as in \citetalias{bennet_patel_2025} to describe satellite galaxies, namely:

\begin{itemize}
    \item satellite: a galaxy located within the virial radius of a more massive galaxy's halo.
    \item bound satellite: a galaxy located within the virial radius of a more massive galaxy's halo that has a 3D velocity below the escape velocity at the same distance.
\end{itemize}

While our definition for a satellite is rather broad, it is chosen to be consistent with observational surveys \citep[e.g., SAGA and ELVES;][]{Mao_2024, Carlsten_2022}. It is worth noting that in the context of satellites around satellites, especially those hosted by a more massive halo, like M33 in M31's halo, varying definitions have been used in the literature \citep[see also][]{Patel_2020, Nadler_2020, Erkal_2020, CorreaMagnus_2022,Battaglia_2022}. %Our adopted definitions are not selected to constrain whether satellites have been long-term companions of their host galaxies, but similar terminology has been used to imply such \citep[e.g.,][]{Patel_2020, Battaglia_2022}.

\subsection{Orbital Uncertainties}
\label{subsec:uncertainties}
Uncertainties on the orbital histories are estimated by computing 1,000 orbits for each of the two dwarfs, initializing each orbit with a 6D phase space vector drawn from the joint measurement uncertainties on the distances, PMs, and LOS velocity measurements in a Monte Carlo fashion. Figure \ref{fig:orbit_errors} shows the corresponding orbital uncertainties as gray lines in the top and middle rows. These uncertainties include the corresponding phase space uncertainties for M33, M31, and the LMC. All galaxy masses remain fixed across the propagation of fiducial orbits. Summary statistics corresponding to uncertainties on orbital parameters are compiled in Table \ref{tab:orbit_params_m33} and will be discussed in Section \ref{sec:orbits}. 

\section{Results: Orbital Histories}
\label{sec:orbits}
\subsection{Direct Orbital Histories}
\label{subsec:direct}
\begin{figure*}[ht!]
\centering
    \includegraphics[width=0.65\textwidth, trim=18mm 0mm 0mm 0mm]{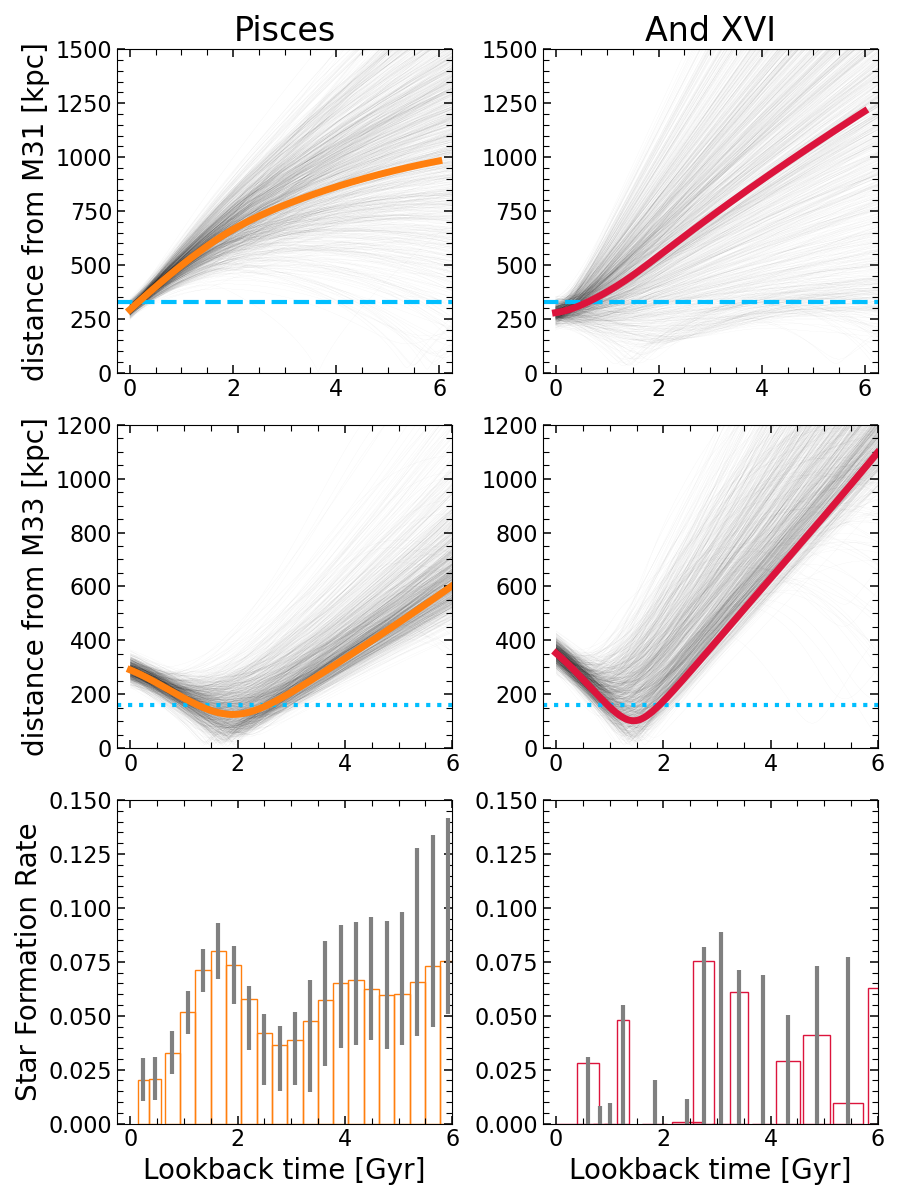}
    \caption{\textbf{Top:} Direct orbital histories and orbital uncertainties between Pisces (left) and And~XVI (right) relative to M31. The \textbf{blue} dashed, horizontal line is the virial radius of M31. \textbf{Middle:} Same as the top row, but orbits are shown relative to M33. The dotted horizontal line is the virial radius of M33. \textbf{Bottom:} The measured SFH of Pisces from \citet{Hidalgo_2011} and from \citet{Savino_2025} for And~XVI. For Pisces, there is a clear correlation between the time of pericenter around M33 (1.91 Gyr ago) and a burst of star formation. For And~XVI, the SFH uncertainties lead to less conclusive evidence of correlations between the orbital and star formation histories.}
    \label{fig:orbit_errors}
\end{figure*}

The direct orbital histories derived by integrating the 6D phase space information backward in the orbital framework described in Section \ref{subsec:orbit_models} are shown in the top two rows of Figure \ref{fig:orbit_errors}. The top row corresponds to orbits for Pisces and And~XVI relative to M31 for the fiducial M33 and M31 masses. Both galaxies crossed into M31's virial radius ($R_{vir}=329$ kpc, dashed green line) within the last $\sim$1 Gyr and are on first infall. As such, they are present-day M31 satellites.

The middle row of Figure \ref{fig:orbit_errors} shows the orbits of both dwarfs relative to M33, which is a satellite of M31 itself \citep[see][]{Patel_2017a, Patel_2025}. While both dwarfs had a recent pericentric passage within M33's virial radius ($R_{vir}=161$ kpc, dotted green line), they are moving away from M33 and are outside its virial radius. Therefore, neither dwarf is presently a satellite of M33.

Figure \ref{fig:orbit_cross} illustrates the direct orbital histories for Pisces and And~XVI as projections in each plane of the M31-centric reference frame. Here, the $z$-axis is normal to M31's disk, which lies in the $x-y$ plane. Dashed circles represent the virial extents of M31 and M33. The orbits of Pisces and And~XVI are primarily confined to the $x-z$ plane, as illustrated in the right-most panel of Figure \ref{fig:orbit_cross}. 

A shared orbital plane implies a similar orbital pole (i.e., direction of the angular momentum vectors), which is typical in the case of group accretion, as seen in the satellites of the LMC \citep{Sales_2017, Jahn_2019}. However, sharing an orbital pole does not strictly correlate with group infall, as several MW satellites (i.e., Carina and Fornax) have been shown to have similar orbital poles to the LMC and its satellites, even if they are not thought to have been accreted with the LMC \citep{Patel_2020}. Additionally, alternative scenarios for similar orbital planes and poles are also possible, such as accretion along the same cosmic filaments \citep[e.g.,][]{Libeskind_2011}, or some combination of these two possibilities. We will further discuss the orbital poles of Pisces and And~XVI in Section \ref{sec:discussion}.

\subsection{Orbits Relative to M31}

In \citetalias{bennet_patel_2025}, Pisces is concluded to be on its first infall into M31's halo. Statistics in the top of Table \ref{tab:orbit_params_m33} show that few Pisces orbits contain a pericenter (2\%) and/or apocenter (18\%). When an apocenter does occur, Pisces orbits M31 at large distances ($\gtrsim$ 500 kpc). 

And~XVI is on its first passage around M31. It is more likely to have completed a pericenter (28\%) within M31's halo, hence the distinction between first passage and first infall (i.e., no previous, close passage). For 7\% of orbits, apocenters are also recovered, typically at distances within M31's virial radius. Nevertheless, the likelihood of both dwarfs never interacting closely with M31 is high ($\gtrsim$ 70\%), unless the adopted mass for M31 is doubled (see Section \ref{subsec:mass_prop_results}).

\begin{deluxetable*}{lcccccc}[ht!]
%\tablenum{3}
\tablecaption{Orbital Parameters for Pisces and And~XVI\label{tab:orbit_params_m33}}
\tablewidth{0pt}
\tablehead{
\colhead{Dwarf}  &  \colhead{$\rm f_{peri}$} & \colhead{$\rm t_{peri}$} & \colhead{$\rm r_{peri}$} &  \colhead{$\rm f_{apo}$} & \colhead{$\rm t_{apo}$} & \colhead{$\rm r_{apo}$} \\
\multicolumn1c{} & \colhead{(\%)} &  \colhead{(Gyr)} & \colhead{(kpc)} &  \multicolumn1c{(\%)} & \multicolumn1c{(Gyr)}  & \multicolumn1c{(kpc)} 
}
%\decimalcolnumbers
\startdata 
\multicolumn{7}{c}{\textbf{Relative to M31 (Fiducial Masses)}} \\ \hline
Pisces & 2 & $\cdots$ [4.93, 5.63, 5.86] & $\cdots$ [32, 95, 572] & 18 & $\cdots$ [2.26, 3.86, 5.22] & $\cdots$ [508, 609, 718] \\ 
And~XVI & 28 & $\cdots$ [0.09, 0.67, 1.86] & $\cdots$ [158, 256, 294] & 7 & $\cdots$ [4.22, 5.10, 5.65] & $\cdots$ [277, 311, 359] \\ \hline 
\multicolumn{7}{c}{\textbf{Relative to M33 (Fiducial Masses)}}  \\ \hline
Pisces & 100 & 1.91 [1.37, 1.75, 2.01] & 125 [67, 140, 212] & 2 & $\cdots$ [4.98, 5.38, 5.9] & $\cdots$ [461, 585, 741] \\ 
And~XVI & 100 & 1.44 [0.82, 1.18, 1.42] & 101 [92, 183, 269] & 7 & $\cdots$ [4.01, 5.19, 5.79] & $\cdots$ [667, 888, 1182]  \\ \hline
\multicolumn{7}{c}{\textbf{Relative to Each Other (Fiducial Masses)}}  \\ \hline
 & 72 & 0.67 [0.33, 0.68, 0.91] & 116 [54, 104, 132] & 13 & $\cdots$ [1.51, 1.89, 5.14] & $\cdots$ [144, 239, 832]\\  \hline \hline 
\multicolumn{7}{c}{\textbf{Relative to M31 (All Masses)}} \\ \hline
Pisces & 7 & [3.58, 4.92, 5.64] & [46, 117, 520] & 27 & [1.74, 3.02, 4.88] & [488, 588, 711] \\
And~XVI & 28 & [0.09, 0.85, 1.88] & [161, 255, 294] & 10 & [3.56, 4.69, 5.53] & [273, 320, 379] \\ \hline 
\multicolumn{7}{c}{\textbf{Relative to M33 (All Masses)}}  \\ \hline
Pisces & 100 & [1.36, 1.71, 2.02] & [61, 136, 209] & 8 & [4.04, 4.95, 5.64] & [394, 518, 763] \\ 
And~XVI & 100 & [0.83, 1.17, 1.41] & [89, 181, 267] & 12 & [3.60, 4.64, 5.55] & [613, 805, 1048]\\ \hline 
\multicolumn{7}{c}{\textbf{Relative to Each Other (All Masses)}}  \\ \hline
& 75 & -- [ 0.34, 0.70,  0.95] & -- [55, 105, 133] & 18 & -- [1.41, 2.08, 4.96 ] & -- [145, 276, 818] \\ \hline
\enddata
\tablecomments{Orbital parameters calculated relative to M31  ($\rm R_{vir}=329$ kpc; top section) M33 ($\rm R_{vir}=161$ kpc; middle section), and for Pisces and And~XVI relative to one another (bottom section) for the fiducial M33 and M31 masses. $\rm f_{peri}$ and $\rm f_{apo}$ denote the fraction of 1,000 orbits where a pericenter or apocenter exists in the last 6 Gyr. Other parameters are taken from the direct orbital histories shown in Figure \ref{fig:orbit_errors}, while values in brackets quote the [15.9, 50, 84.1] percentiles calculated from the set of 1,000 orbits computed for each galaxy (gray lines in the top and middle row of Figure \ref{fig:orbit_errors}). The bottom half of the table shows the orbital parameters, now taking into account the orbital uncertainties across all nine explored mass ranges for M33 and M31. As the statistics summarize nine mass combinations, we do not report central values from any particular set of direct orbital histories. We refer readers to Figures \ref{fig:mass_combo1},\ref{fig:mass_combo2},\ref{fig:mass_combo3} for direct orbital histories across all mass combinations.}
\end{deluxetable*}

\subsection{Orbits Relative to M33: Were Pisces and And~XVI Once M33 Satellites?}
\label{subsec:m33sats}
\subsubsection{Pisces} 
\label{subsec:pisces}

There has been much discussion historically about whether Pisces is a member of the M31 or M33 system \citep[][]{Lee_1995,Aparicio_1997,Cook_1999,McConnachie_2005,McConnachie_2012}. Recent RR Lyrae distance estimates \citep[][]{Savino_2022} move Pisces significantly closer to the MW compared to previous TRGB estimates \citep[][]{McConnachie_2012}. This places Pisces at 293$\pm$6 kpc from M31, just within the virial radius of M31 ($R_{vir}=329$ kpc). The new 3D separation between Pisces and M31 is thus almost identical to that between Pisces and M33 (290~kpc). The orbits presented in this work allow a more thorough investigation of how Pisces has evolved in the M31-M33 system.

\begin{figure*}
\centering
    \includegraphics[width=0.99\textwidth, trim=18mm 5mm 0mm 0mm]{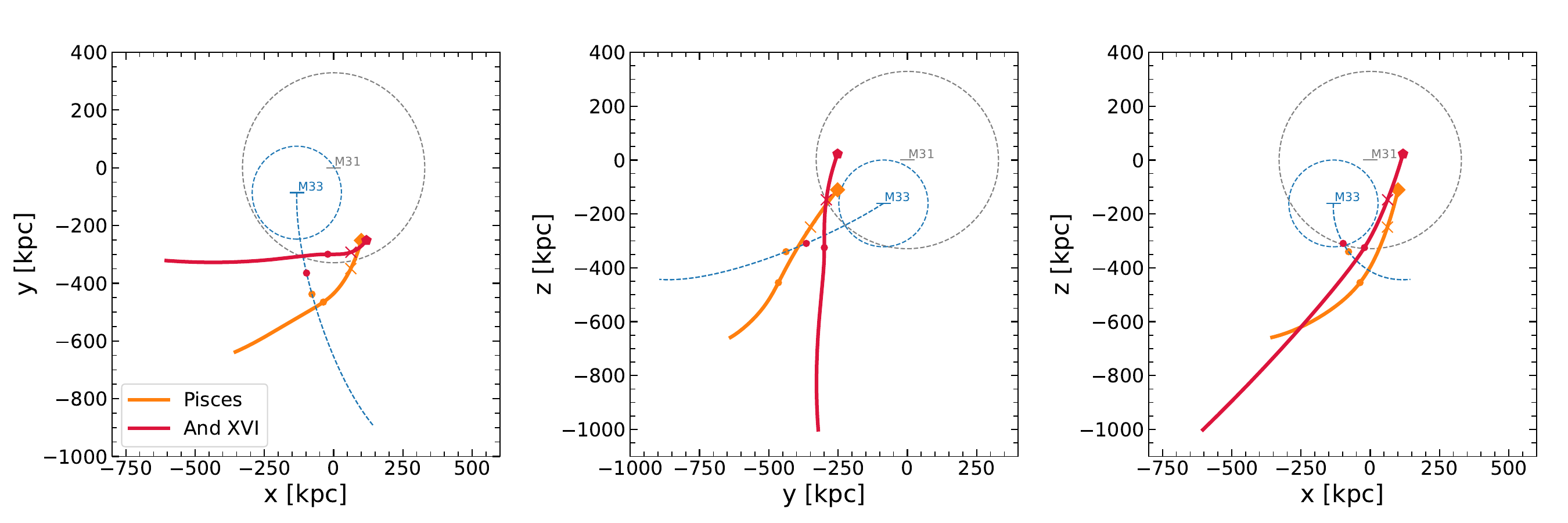}
        \includegraphics[width=0.99\textwidth, trim=18mm 10mm 0mm 0mm]{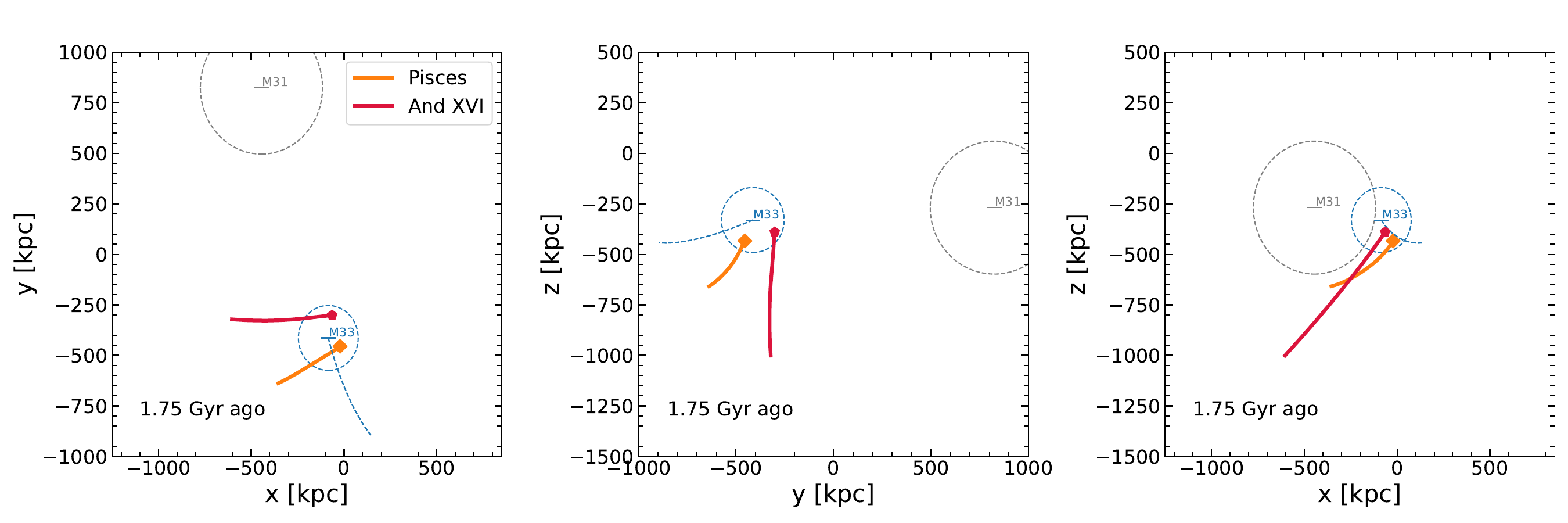}
    \caption{\textbf{Top:} Spatial projection of the direct orbital histories provided in Figure \ref{fig:orbit_errors} in the M31-centric frame\footnote{This reference frame is the Galactocentric frame, translated to the location of M31. This does not account for the geometry (position angle and inclination) of M31 as in \citet{Patel_2025}.} over the last 6 Gyr. Markers indicate the positions of the dwarfs today. The virial extent of M31 (gray) and M33's (blue) halos is indicated by dashed circles. Pisces and And~XVI are within M31's halo but not within M33's halo at present; however, both dwarfs have crossed through M33's halo in the past. Small closed circles show the time of pericenter relative to M33, while the "x" symbols indicate the timing of pericenter between the dwarfs. Similarly, small closed circles on M33's trajectory correlate to the timing of the dwarfs' pericenter around M33. \textbf{Bottom}: Same spatial projection but shown between 1.75 Gyr ago, the time splitting when both dwarfs have a pericenter around M33, and 6 Gyr ago. Both dwarfs clearly reside inside the halo of M33 at the time of pericenter. }
    \label{fig:orbit_cross}
\end{figure*}

\begin{figure*}[ht!]
\centering
    \includegraphics[width=0.4\textwidth, trim=18mm 10mm 0mm 0mm]{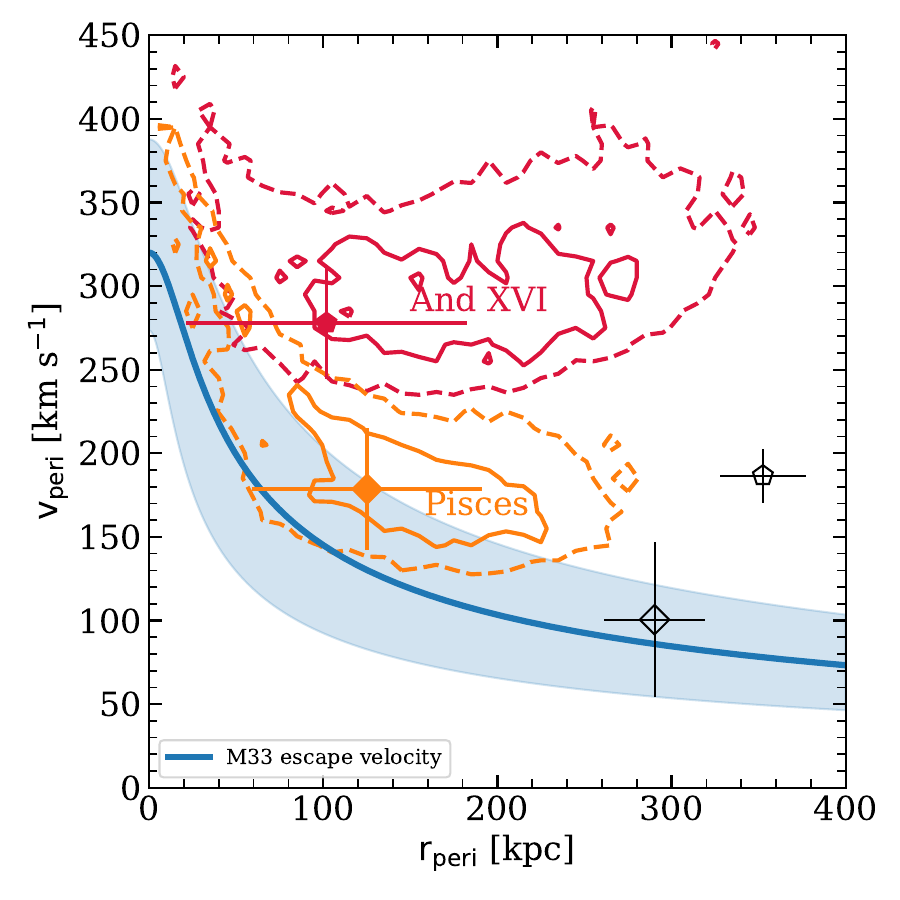}
    \includegraphics[width=0.4\textwidth,trim=0mm 10mm 18mm 0mm]{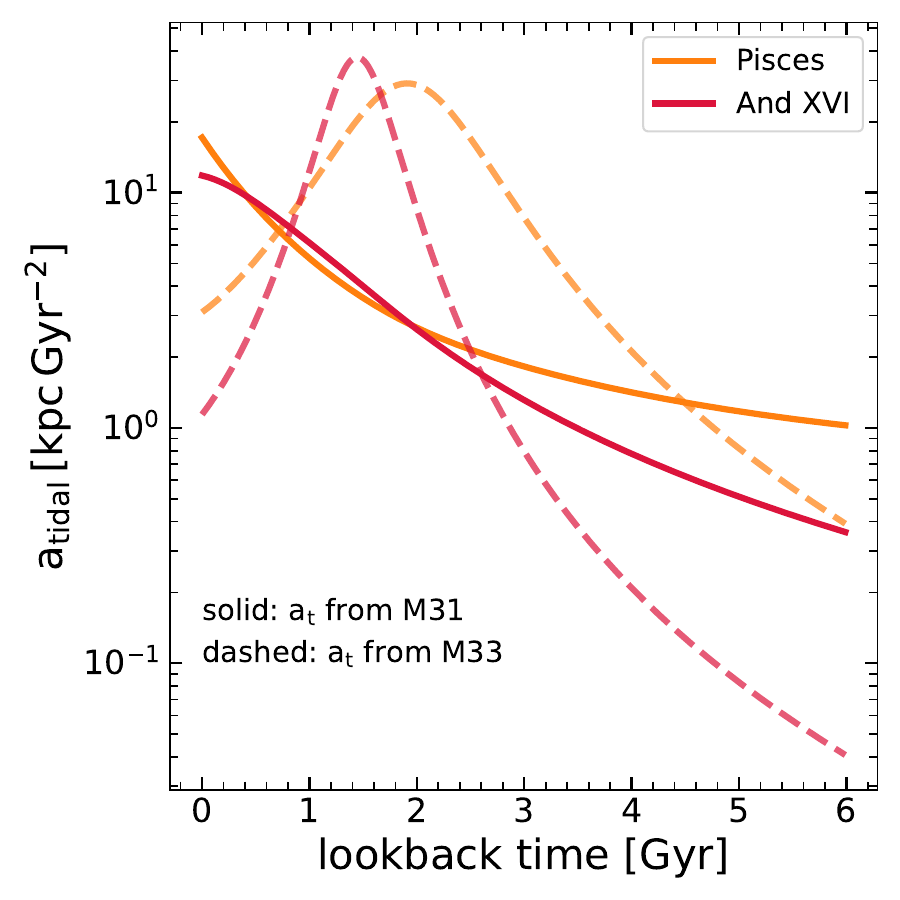}
        \caption{\textbf{Left:} Distance and velocity of both dwarfs with respect to M33 at the time of pericenter. Error bars represent the standard deviations of these quantities from the 1,000 orbit calculations sampling the 6D phase space information measurements in the fiducial mass combination. Open symbols indicate the position and velocity of both dwarfs at the present day. The solid blue curve illustrates M33's escape velocity as a function of distance for the fiducial M33 mass, while the shaded region indicates the escape velocity curve extents for the lower and higher M33 masses. The 1$\sigma$ and 2$\sigma$ contours are also shown for the combined propagation of orbital uncertainties across all mass combinations. Both Pisces and And~XVI were previously M33 satellites and have a high likelihood of experiencing a fly-by encounter with M33 in the past. \textbf{Right:} The tidal acceleration, $a_{tidal}$, exerted on Pisces and And~XVI from M31 (solid lines) and M33 (dashed lines) as a function of time in the fiducial setups. $a_{tidal}$ is computed between the position of each dwarf relative to M31/M33 as a function of time, compared to tidal acceleration at the virial radius of each dwarf ($R_{vir}=$44 kpc for Pisces and $R_{vir}=$26 kpc for And ~XVI). The tidal acceleration from M33 is stronger than from M31 for both dwarfs, until the last $\sim$1 Gyr when M31's tidal acceleration becomes more dominant.}
    \label{fig:M33_esc}

\end{figure*}

 In the fiducial mass setup, Pisces completed a pericenter around M33 at 1.91 Gyr ago, just 125 kpc from M33. This occurrence is statistically common across the 1,000 orbit calculations discussed in \citetalias{bennet_patel_2025}, such that 100\% of Pisces orbits have a pericenter with respect to M33 at $\sim$1.4-2 Gyr ago at distances of $\sim$67-212 kpc, where these ranges represent the 1$\sigma$ credible intervals (see Table \ref{tab:orbit_params_m33}). For 60.5\% of orbits, the pericenter occurs within the virial radius of M33, making Pisces a highly likely past satellite of M33. The bottom panel of Figure \ref{fig:orbit_cross} shows the spatial projections of Pisces' orbit at 1.75 Gyr ago, when it was inside M33's virial radius and much farther from M31 than it is today. 

The distance and velocity of Pisces at pericenter relative to M33 are illustrated as the orange diamond in the left panel of Figure \ref{fig:M33_esc}. Errors correspond to the standard deviation of position and velocity at pericenter. Also shown is M33's escape velocity curve (blue line) derived assuming the fiducial M33 mass. This panel shows that while Pisces was previously nearby (125 kpc) to M33 during its most recent pericentric passage, its velocity at pericenter was $\sim$175 km s$^{-1}$ whereas the escape speed for M33's halo at the distance of Pisces is just $\sim$130 km s$^{-1}$. This quick passage implies that Pisces did not evolve in the same environment as M33 for any substantial period of time, as is likely the case for the LMC and its associated group of satellites, at least some of which have been shown to remain bound in a group configuration and therefore have evolved together for some time.

The open diamond in the left panel of Figure \ref{fig:M33_esc} shows Pisces's 3D position and velocity relative to M33 today. These indicate that while there is a chance that Pisces is bound to M33 today based on its velocity uncertainties and corresponding intersection with M33's escape speed curve, it is currently significantly outside the virial radius of M33 ($R_{vir}$=161 kpc), and would therefore not be counted as an M33 satellite today.

We conclude that while Pisces was previously a satellite of M33, it was not bound to M33 at the time of its passage; instead, it had a past fly-by encounter with M33. Despite a high-speed passage around M33, M33's gravitational influence dominates over M31's until the last $\sim$1 Gyr, when Pisces begins to approach M31 for the first time. The right panel of Figure \ref{fig:M33_esc} shows the tidal acceleration exerted by M33 and M31 along the orbit of Pisces relative to the tidal acceleration at Pisces's virial radius. At the time of Pisces' pericenter around M33, it was concurrently well outside the virial radius of M31. 

The interaction between Pisces and M33 coincides with a spike in the star formation activity in Pisces \citep[][]{Hidalgo_2011}, which begins 2.5 Gyr ago and peaks 1.5 Gyr ago (see bottom left panel of Figure \ref{fig:orbit_errors}). The distance of Pisces at pericenter is such that gas compression could be strong enough to yield this burst. In a small percentage of orbits (0.4\%), Pisces has an earlier pericenter passage around M33 between $\sim$5-6 Gyr ago at distances ranging from 87-535 kpc. However, even if this pericentric passage did occur, it does not correlate with any measured bursts in Pisces' SFH.

Interestingly, Pisces is also known to be close (30 arcmin, 5.2 kpc at Pisces distance) to an HI cloud. This HI cloud has a line-of-sight velocity relative to Pisces of 45 km s$^{-1}$. This similar position and velocity combination is unusual and would only be expected by coincidence 4\% of the time \citep{Robishaw_2002}. 

One possibility is that the gas cloud was expelled from Pisces during the past interaction with M33 via ram pressure or tidal stripping. However, the cloud's line-of-sight velocity multiplied by the time of Pisces' pericenter around M33 (1.9 Gyr) yields 87 kpc; thus, the distance between the cloud and Pisces was greater in the past. 

If the cloud were expelled from Pisces, this would increase the original gas fraction for Pisces from $M_{HI}/M_{*}\sim0.2$ as reported in \citet[][]{McConnachie_2012, Putman_2021} to $M_{HI}/M_{*}\sim1.85$ before this potential gas stripping. The latter is far more similar to field dwarfs or other gas-rich dwarfs in the LG, which tend to have gas fractions between 1 and 10 \citep[][]{Robishaw_2002,Putman_2021}. 

A coincidental alignment is possible since this region of the sky has significantly more HI clouds compared to the LG as a whole \citep[][]{Westmeier_2007}. Alternatively, the cloud could have been dislodged from Pisces by M31, given that ram pressure from M31 is stronger than that from M33; however, tidal stripping is unlikely since M33's tidal acceleration is much larger than that from M31 at the time of pericenter (see right panel of Figure \ref{fig:M33_esc}). Thus, while the 3D velocity vector suggests a tentative association between the gas cloud and Pisces, it is not conclusive without a firm distance measurement.

\subsubsection{And~XVI}
\label{subsec:and16}
And~XVI is presently 280 kpc from M31 and 352 kpc from M33. While there has been little to no speculation of And~XVI's association with M33, our orbits revealed the possibility of such a scenario, warranting further analysis.

The right middle panel of Figure \ref{fig:orbit_errors} shows the orbit of And~XVI relative to M33, which indicates And~XVI completed a pericenter around M33 at a distance of 101 kpc at 1.44 Gyr ago according to the direct orbital history. And ~XVI completes a pericenter around M33 in all 100\% of orbits, albeit within an extensive range of possible distances compared to Pisces (i.e., $\sim$92-269 kpc; see Table \ref{tab:orbit_params_m33}). In 41.6\% of orbits, And~XVI reaches pericentric distances within M33's virial radius, and is therefore characterized as a previous satellite of M33. The bottom panels of Figure \ref{fig:orbit_cross} show the spatial projection of And~XVI's orbit at 1.75 Gyr ago when it was just past pericenter around M33.

The red pentagon in Figure \ref{fig:M33_esc} shows the distance and velocity of And~XVI at pericenter with respect to M33. Like Pisces, And~XVI completed a pericentric passage about M33 while moving at a high speed ($\sim$275 km s$^{-1}$), far exceeding the escape speed at that distance ($\sim$140 km s$^{-1}$). The right panel of Figure \ref{fig:M33_esc} indicates that M33's tidal acceleration on And~XVI was stronger than that of M31 beyond $\sim$1 Gyr ago, but M33's gravitational influence dissipates as And~XVI retreats from M33 and approaches M31. Despite a strong tidal acceleration, And~XVI is unlikely to have evolved in the same environment as M33 for any substantial period of time, given the high-speed encounter. But, as shown in the right panel of Figure \ref{fig:M33_esc}, M33 exerts a significant tidal force on the dwarfs (1-4 Gyr ago for Pisces and 1-3 Gyr ago for And~XVI)\footnote{Despite a significant tidal force from M33, neither And~XVI nor Pisces pass within the tidal radius (16 kpc and 22 kpc, respectively, for And~XVI and Pisces in the fiducial mass model), calculated using Equation 3 from \citet{vandenBosch_2018}, and are therefore unlikely to experience tidal stripping owing to M33.}. This is likely to have altered the trajectories of both dwarfs, making them co-planar within M31's halo and subsequently leading to their passage around each other.

The open pentagon in the left panel of Figure \ref{fig:M33_esc} shows And~XVI's present 3D velocity and position relative to M33. These values show that And~XVI is not an M33 satellite today. We conclude that, like Pisces, And~XVI was previously a satellite of M33, but it was not bound to M33 in the past. Additionally, it was not a bound satellite of M31 at the time of its interaction with M33 (see top right panel of Figure \ref{fig:orbit_errors}).

The SFH of And~XVI from \citet{Savino_2025} is illustrated in the bottom right panel of Figure \ref{fig:orbit_errors}. And~XVI's SFH shows several features, which set it apart from other M31 system UFDs. Firstly, it seems to have produced most of its stars after reionization and then continued to form stars, albeit at a reduced rate, until final quenching occurs relatively recently ($\tau_{90}=7.8^{+0.5}_{-0.4}$ Gyr ago), but before its interaction with M33. While there is a high uncertainty on measurements of the recent SFH of And~XVI in particular, it appears to have quenched much later than most other ultra-faint dwarfs in the halo of M31, which form the majority of their stellar mass before reionization, even if some (e.g., And~XXII) show extended star formation at intermediate times \citep{Savino_2025}. 

Thus far, the interactions between each dwarf and the M31-M33 system have been discussed as separate instances; however, it is also possible that Pisces and And~XVI are dynamically associated with one another. Section \ref{sec:discussion} will discuss this possibility in more detail.

\subsection{Exploring Different Combinations of M33 and M31 Masses}
\label{subsec:mass_prop_results}

Many recent studies have shown that the mass of M31 remains uncertain, with measurements spanning $1-3 \times 10^{12}\, M_{\odot}$ \citep[see][for a recent compilation]{bhattacharya25}. Thus far, we have considered an M31 virial mass exactly in the middle of this range at $2 \times 10^{12}\, M_{\odot}$. More importantly, the mass ratio between M31 and M33 primarily sets the orbital family of solutions that these two galaxies are most likely to have when using backward orbit integration, as shown in our previous works \citep{patel17a, Patel_2025}.

As the dynamics of the M31-M33 system, and subsequently any less massive counterparts, depend on the assumed masses of M31 and M33, here, we explore a combination of different masses for both galaxies to determine whether our conclusions are unique to the assumptions of our fiducial model or not. 

For M31, we consider masses of $(1.5, 2, 3) \times 10^{12}\, M_{\odot}$ and for M33, we consider $(1, 2.5, 5) \times 10^{11}\, M_{\odot}$. These masses are consistent with those determined from abundance matching \citep[e.g.,][]{moster13}. We repeat the methods used to integrate direct orbital histories, as well as the corresponding 1,000 orbital solutions spanning the uncertainties on 6D phase space information for M31, M33, the LMC, and the dwarf galaxies. Thus, all nine mass combinations result in a combined set of 9,000 orbital solutions that span a uniform prior across the adopted masses. Using these orbital solutions, we report summary statistics on orbital parameters in the bottom half of Table \ref{tab:orbit_params_m33} (see results for ``All Masses" vs. ``Fiducial Masses"). Direct orbits and orbital uncertainties for each of the nine mass combinations are shown in Figures \ref{fig:mass_combo1}, \ref{fig:mass_combo2}, and \ref{fig:mass_combo3}.

In this set of figures, the orbits of both dwarfs relative to M31 are shown in the first and third columns, respectively, while orbits relative to M33 are shown in the second and fourth columns. Each row corresponds to a different M33 mass, with M33 mass increasing from top to bottom. Printed on each panel are the statistics corresponding to the distance at pericenter ($r_{peri}$) and time at pericenter ($t_{peri}$), such that numbers in the square brackets correspond to the [15.9, 84.1] percentiles. $f_{peri}$, the percentage of 1,000 orbits for a specific mass combination where a pericenter exists, is also printed on each panel. For the panels showing orbits relative to M33, we instead list the fraction of pericenters with distances less than the virial radius of M33. Below, we highlight the main dynamical conclusions drawn from varying the M31-M33 mass combinations.

To begin with, we examine how conclusions change when the mass of M31 is varied. For the lowest mass M31, both Pisces and And~XVI remain on first passage for a statistical majority of orbits. Moving to the intermediate (Fig. \ref{fig:mass_combo2}) and high mass (Fig. \ref{fig:mass_combo3}) M31, an increasing fraction of orbits have a pericentric passage as the gravitational influence of M31 (and M33) increases, particularly for Pisces. For these higher M31 mass scenarios, Pisces experiences a pericenter with respect to M31 at earlier times compared to the lowest mass M31, but the changes for And~XVI are less stark. The pericenter time remains between $\sim$0-2 Gyr ago for And~XVI. Thus, Pisces is more susceptible to the mass of M31; all three orbital parameters corresponding to pericenter vary significantly as the masses of M33 and M31 are varied. In contrast, the And~XVI orbital parameters computed relative to M31 are nearly agnostic to different mass combinations.

The likelihood that either dwarf has never interacted closely with M31 decreases with increasing mass. However, even in high mass M31 models, a substantial fraction of orbits still indicate recent infall, not long-term companionship. We conclude that massive M31 models weaken but do not overturn the first infall interpretation.

Across all explored M31–M33 mass combinations, both Pisces and And XVI experience a pericentric passage relative to M33 in 100\% of orbits. Both the timing and distance at pericenter relative to M33 are not driven by the choice of mass. However, the adopted M33 mass does affect how likely it is that the encounter with M33 occurred within its virial radius. A more massive M33 has a larger virial radius and higher escape speed as a function of distance from M33, thus a higher percentage of orbits are classified as prior M33 satellites, especially in the case of Pisces. Orbital histories shown in Appendix \ref{sec:appendix} illustrate that the percentage of Pisces' orbits with a pericenter inside of M33's virial radius increases by approximately 20\% (from 50-55\% to 70-75\%) with increasing M33 mass if M31's mass is held fixed. And XVI also shows an increased percentage, but only at a 5-6\% level.

The contours in Figure \ref{fig:M33_esc} illustrate the $1\sigma$ and $2\sigma$ contours for the orbital parameters with respect to M33 across all mass combinations, while the blue shaded region corresponds to the escape velocity curve spanning the lowest to highest mass M33. The fiducial values (closed symbols) lie primarily in the $1\sigma$ contours. Figure \ref{fig:M33_esc} shows the Pisces contours and the blue shaded region denoting M33's escape velocity overlap significantly, whereas the And XVI contours and error bars barely do.

Together, these results show that both M31-centric and M33-centric orbital histories are mass-dependent, but in different ways. M31's mass modulates the timing of pericenter, whereas M33's mass most commonly alters the distance reached at pericenter relative to M33. The fraction of orbits where pericenter occurs within M33’s virial radius depends strongly on M33’s mass, increasing for higher mass M33 models and decreasing for lower mass ones. Fly-by encounters are seen in nearly all cases, even for the most massive halos considered. Varying the M31-M33 masses primarily modulates the likelihood of prior satellite status, while preserving the conclusion that both Pisces and And XVI underwent recent, rapid interactions with M33. Overall, Pisces is more substantially affected by varying the masses of M31 and M33 compared to And XVI.

\section{Discussion}
\label{sec:discussion}
\subsection{Quantifying the Possibility of Group Infall and Satellite Exchange}

\subsubsection{Group Infall}
\label{subsubsec:group}

Groups of dwarfs accreted into the halos of more massive galaxies have been particularly interesting recently, given the presence of LMC satellites that were accreted along with it onto the MW's halo. Furthermore, the properties (i.e, quenching times and r-process enrichment) of these LMC-associated dwarfs also suggest they evolved together before group accretion into the MW \citep{Sacchi_2021, Durbin_2025, Ji_2023}. Studies on the orbits of the Magellanic Clouds and Magellanic satellites have shown that in specific models, even when the Clouds or corresponding satellites become unbound, they continue to orbit in the same orbital plane, retaining closely aligned angular momenta long after infall, leading to interactions as a consequence of sharing an orbital plane \citep[e.g.,][]{Li_2008, Nichols_2012,Pardy_2019,Vasiliev_2024}. The shared orbital plane of Pisces and And~XVI is reminiscent of a similar scenario, where Pisces and And~XVI could have been bound to another galaxy (M31, M33, each other, or another galaxy), perhaps beyond the 6 Gyr time period explored in this work.

The similarity of pericenter timing around M33 lends evidence to a link between the two dwarfs. To determine how often the passage of both dwarfs around M33 is coincident in time, we first fit a normal distribution to the distribution of time at pericenter relative to M33, $t_{peri}$, for each dwarf. Using these normal distributions, we find the Overlapping Coefficient (OC), a metric that measures the similarity between probability distributions by computing the area of intersection between their density functions. Overlapping Coefficients are typically reported on a scale of 0 to 1, where 1 implies 100\% intersection. Following this procedure, the OC for $t_{peri}$ between Pisces and And~XVI is 0.0037; in other words, 0.37\% of orbits show the dwarfs orbiting around M33 simultaneously\footnote{Across all mass combinations, the OC ranges between 0-3.8\%.}. However, dwarfs can also orbit the host galaxy (M33 in this case) at varying but similar times, even in a group scenario, as is the case for satellites of the LMC. For example, if we use the 1,000 orbital uncertainties to compare how often $t_{peri}$ for Pisces is $\pm$0.25 Gyr (0.5 Gyr) of $t_{peri}$ for And~XVI, the frequency increases to 28\% (51\%). 

Noting the similarities in the distance and timing of pericenter between Pisces-M33 and And~XVI-M33, we also investigated whether these two dwarfs could be dynamically associated with one another, lending further evidence to the possibility of a previous group accretion onto either M33, M31, or both. In Table \ref{tab:orbit_params_m33}, we report summary statistics for the orbital parameters calculated from the relative orbit between Pisces and And~XVI. Note, this does not account for any gravitational influence that each dwarf may exert on the other; rather, these statistics summarize the relative difference between the 1,000 orbits computed for each dwarf galaxy (including all five bodies) independently. 

In more than 70\% of orbits, the dwarfs complete a pericentric passage around each other, typically in the last 1 Gyr, after they have each passed around M33. The distances at pericenter are $r_{peri}=116^{+16}_{-62}$. These conclusions are robust across all explored mass combinations. While this is not a very close encounter for dwarfs of this mass range, it too suggests that And~XVI and Pisces could be forming or were previously a loosely associated pair, or even part of a group accretion into the M31 (or M33) halo. However, with the current precision of 6D phase space information and the limitations of rigid orbital models, it is not possible to constrain whether the dwarfs were accreted together or as separate accretion events onto either M33 or M31.

Model limitations also include the fact that backward orbit integration tends to underestimate group configurations, and that our orbital methods do not account for the gravitational influence of the dwarfs on each other. Cosmological zoom-in simulations of dwarfs in environments analogous to the M31-M33 system may reveal helpful insight into the origin of dwarfs with orbits similar to And~XVI and Pisces, but this is beyond the scope of the present work. 

%In Section \ref{subsec:m33sats}, we discussed that the timing of pericenter relative to M33 is typically more recent for And~XVI compared to Pisces (see also Table \ref{tab:orbit_params_m33}). However, we find that for 9.9\% of orbits, Pisces orbits M33 more recently.  
%Disentangling these possible scenarios with reconstructed orbits will require improved PM measurements for the dwarfs and M31. 

\begin{figure*}
\centering
    \includegraphics[width=0.8\textwidth, trim=18mm 0mm 18mm 0mm]{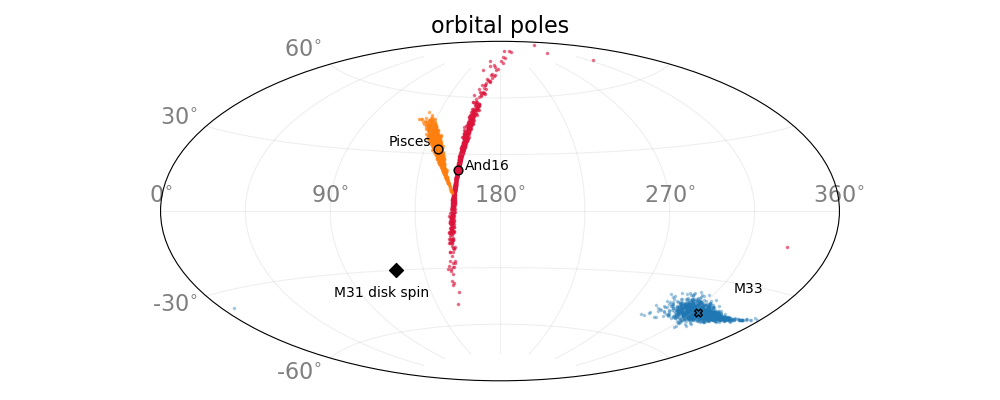}
    \caption{Orbital poles of Pisces and And~XVI relative to M31 in Galactic coordinates. Also shown is the orbital pole of M33 relative to M31 and the M31 disk spin. Open symbols show the orbital poles computed with the coordinates in Table \ref{tab:Pos_vel}. Closed symbols represent orbital poles calculated using the 1,000 samples drawn from the Monte Carlo scheme used to propagate uncertainties on observational quantities used to derive 6D phase space information. See Section \ref{subsec:uncertainties} for more details.}
    \label{fig:orbpoles}
\end{figure*}

\subsubsection{Satellite Exchange}
\label{subsubsec:exchange}
In most orbits, Pisces and And~XVI are on their first infall into M33, experiencing a single orbital pericenter. However, in some subset of orbits, the dwarfs may have been M31 satellites already (i.e., within the virial radius of M31) before passing around M33 and finally transferring back to M31. This exchange of dwarfs between the two systems appears in only 2.7\% of orbits for Pisces and 0.3\% for And~XVI in the fiducial mass scenario. This possible transfer of dwarfs is intriguing, as it can bias comparisons between observational surveys, which assume static galaxy positions, and dynamical analyses like these, which follow the center of mass evolution of galaxies over time. More theoretical work is needed to understand further the frequency of satellite exchange in a cosmological framework.

\subsubsection{The Probability of Independent Accretion Events}
Thus far, we have examined the possible interaction scenarios between Pisces, And~XVI, M33, and M31, to find evidence for a dynamically associated pair or group of galaxies. However, it is not impossible that Pisces and And~XVI are two unrelated galaxies that independently passed around M33 and both landed back in the halo of M31 today.

Figure \ref{fig:orbpoles} illustrates the orbital poles, or the direction of angular momentum\footnote{The probability of these dwarfs belonging to the Great Plane of Andromeda (GPoA) \citep[e.g.,][]{Conn_2013, Ibata_2013,SantosSantos_2020}, a proposed spatial and kinematic alignment for about 50\% of M31's satellites will be discussed in Sohn et al., in prep.}, in Galactic coordinates but in an M31-centric frame for Pisces, And~XVI, M31, and M33. The dots around each galaxy indicate the uncertainties on the orbital pole, calculated from the same 1,000 6D phase space vectors used to initialize the orbital uncertainties described in Section \ref{subsec:uncertainties}.

To compute the likelihood that two dwarfs randomly have such similar orbital poles, we calculate the fraction of a sphere's surface area within the cone defined by $\theta$, 
$A=[1-cos(\theta)]/2$, where $\theta$ is the difference in the orbital pole of Pisces and And~XVI. In this case, the orbital pole of Pisces is $(l,b)=(216.7^{\circ}, 32.5^{\circ}$) and that of And~XVI is $(l,b)=(203.5^{\circ}, 21.8^{\circ}$), thus $\theta=17^{\circ}$. This yields a probability of $A$=0.011. Accounting for the 1$\sigma$ uncertainties on $\theta$, this yields probabilities of $A$=0.004-0.013. While this probability calculation is rather simple and the probabilities could be higher if the fact that they interacted with M33 recently were included, it supports that these dwarfs are indeed likely related.

Notably, these dwarfs are not entirely \emph{random}, as they are selected based on characteristics that constrain their position and velocity relative to M33; thus, this null hypothesis is not entirely appropriate. However, only four other M31 satellites (NGC 147, NGC 185, IC 10, M33) of the total thirty-seven have published orbital histories derived from 6D phase space information \citep{Sohn_2020, Patel_2017a, Nidever_2013}, making it challenging to compute more rigorous statistics on the probability of independent versus group accretion with a more well-justified null hypothesis. 

We have computed the statistical frequency by which the dwarfs cross into the virial radius of M33 (see Section \ref{subsec:m33sats}) and how often this occurs simultaneously (see Section \ref{subsubsec:group}), how often the dwarfs are exchanged between the halos of M33 and M31 (Section \ref{subsubsec:exchange}, and the probability that the dwarfs interacted among themselves (Section \ref{subsubsec:group}). All conclusions suggest an association between Pisces and And~XVI despite a more sophisticated proof indicating this association is by chance.

\subsection{Consequences for the M33 Satellite Mass Function}
If Pisces was an M33 satellite in the past and is now an M31 satellite, this raises the question of how to categorize Pisces. The ELVES \citep[][]{Carlsten_2022} survey of dwarf galaxies around MW analogs in the Local Volume has compared the number and properties of dwarfs to those found in the LG, both around the MW and M31. These data, along with other work in the Local Volume, have produced several possible relations between the properties of the host galaxy and the number and properties of the satellites, \citep[e.g.][]{Bennet_2019,Smercina_2022,MP_2024}. However, if Pisces is counted as an M33 dwarf instead, it changes those possible relations, particularly concerning the quenched fraction of satellite systems. 

For instance, if one counts Pisces as an M33 satellite rather than an M31 satellite and examines the M31 system down to the magnitude limit of ELVES ($M_V=-9$), then the star-forming fraction of M31 changes from 17\% (2/12) to 9\% (1/11), a two-fold decrease. This would also increase the star-forming fraction of M33 satellites from zero to one dwarf. This categorization impacts several suggested relationships between satellite number, satellite quenched fraction, host mass, host morphology, largest satellite, host environment, and the most recent accretion event \citep[e.g.][]{Bennet_2019,Javanmardi_2020,Smercina_2022,Muller_2023,Danieli_2023,Karunakaran_2023,MP_2024}. As these relations rely on a relatively small number of Local Volume hosts for which robust satellite system measurements are available, any changes in the M33-M31 system can significantly affect the statistical significance of the proposed relations and attempts to reconcile such ties between observations and simulations. 

On the other hand, if Pisces is a satellite of M33, the revised M33 luminosity function (one classical satellite and three ultra-faint dwarf satellites) is more similar to a typical system of its stellar mass in the Local Volume according to the results of ELVES-DWARFS \citep[][]{Li_2025} or ID-MAGE \citep[][]{Hunter_2025}. Classifying Pisces as an M33 satellite also changes M33 from a 3$\sigma$ outlier to a 2$\sigma$ outlier.
And~XVI is below the magnitude limits for most of the aforementioned Local Volume surveys \citep[$M_{V}=-9$,][]{Carlsten_2022,Li_2025,Hunter_2025}, and therefore it is not discussed in this context. 

Finally, distant M33 satellites may have yet to be discovered, given the limited spatial coverage of previous wide field surveys, which do not examine the entire virial radius of M33 \citep[e.g. Pan-Andromeda Archaeological Survey][]{McConnachie_2009}. Expectations based on simulations show that 8$\pm$4 satellites with $M_* > 10^4\,M_{\odot}$ are expected for an M33-mass galaxy, which includes $\sim$1-3 classical-mass satellites \citep[][]{Dooley_2017,Patel_2018b}. Upcoming data from the Subaru Prime Focus Spectrograph or the Roman Space Telescope may aid in such discoveries, in lieu of a northern counterpart to LSST.

% A more consistent metric for identifying satellites across observations and simulations is needed to characterize these systems appropriately. This example of Pisces and how its host galaxy changes as it interacts with multiple nearby galaxies highlights the limitations of comparing observations without any dynamical information (i.e., no phase space information and therefore no orbits) to those from simulations. 

\subsection{Dwarfs' Impact on M33's Morphology}
The origin of the warps observed in M33's stellar and gaseous disks \citep{McConnachie_2009, Putman_2009} is still debated.
\cite{Corbelli_2024} conclude these warps were not caused by a tidal interaction with M31, as previously proposed. Here, we speculate whether interactions with Pisces and/or And~XVI could have played a role in M33's observed morphology. 

In our direct orbital histories, Pisces completes a pericenter at a distance of 125~kpc from M33. This is likely too distant for Pisces to have caused the M33 disk warps, given the mass ratio between them (1:50 total mass ratio). Similarly, And~XVI only reaches a little closer to M33 (101 kpc) with a mass ratio of 1:250, and is also not expected to be related to M33's warps.

However, in our investigation of phase space uncertainties (see Tables 5 and 6 in \citetalias{bennet_patel_2025} and Table \ref{tab:orbit_params_m33} in this work), the orbits where Pisces approaches M33 the closest have pericenter distances as low as 67 kpc (the lower end of the 1$\sigma$ range). Though this only occurs for a small fraction of orbits, Pisces remains an unlikely but yet possible candidate for M33's warps. On the other hand, the lower 1$\sigma$ range of distances at pericenter for And~XVI is similar to the results of its direct orbital history at 92 kpc, which is likely too distant to cause any morphological changes to M33.

While the outer regions of M33 show a significant star formation rate enhancement at 3-5 Gyrs \citep[][]{Barker_2007}, which could correlate to an interaction or merger with a dwarf galaxy, this timing is before any potential encounter with Pisces or And~XVI based on the orbits presented in this work.

\section{Conclusions}
\label{sec:conclusions}
We have used backward orbital modeling using rigid potentials to reconstruct the orbital histories of Pisces and And~XVI to investigate the dynamical nature of both dwarf galaxies with respect to M33, the most massive satellite in the M31 system, and to each other. M33 is posited to host a population of its own satellite galaxies, like the LMC \citep[e.g.,][]{Patel_2018b}; however, current M33 satellite candidates are yet to be confirmed due to a lack of available 6D phase space information \citep[see][]{Martin_2009, Chapman_2013, Martinez_2022, Collins_2024,Ogami_2024}. 
Both Pisces and And~XVI now have 6D phase space information available, allowing for orbit reconstruction.

In this work, satellite galaxies are defined as those within the virial radius of a more massive host galaxy, akin to observational surveys of satellites around MW/M31-mass and LMC/M33-mass dwarfs \cite[e.g.,][]{Carlsten_2022, Mao_2024, Hargis_2020,Doliva-Dolinsky_2025, Carlin_2024, Davis_2024, Hunter_2025, Li_2025}. As such, both Pisces and And~XVI are presently satellites of M31.

Through our orbit analysis, we find that while Pisces and And~XVI are typically on first infall into the halo of M31 in a majority of sampled orbital histories, both completed a pericentric passage around M33 at distances of $\sim$100-125 kpc in the last $\sim$1-2 Gyr in nearly 100\% of orbits capturing the uncertainties on the measured 6D phase space information. For Pisces, the timing of its pericentric passage around M33 corresponds to a burst in star formation, supporting the physical impact of this interaction. Our results also yield a 28\% chance that both dwarfs orbited M33 together, having pericentric passages within $\pm$0.25 Gyr of each other. While this timing is near-synchronous, the distances and mass ratios involved make it unlikely that either dwarf caused M33's observed disk warps. 

The probability of Pisces and And~XVI interacting with each other is also non-negligible. In more than 70\% of orbits, the dwarfs complete a passage around each other in the last 1 Gyr. While the dwarfs do not interact at very close distances, they orbit in the same plane and have closely aligned orbital poles. This, combined with the similar orbital histories relative to M33 and M31, suggests these dwarfs may be or have been a loosely associated pair or remnants of an earlier group accretion, warranting further investigation. 

We conclude that Pisces and And~XVI were previously satellites of M33 with frequencies of 61\% and 42\%, respectively. However, they were not bound to M33 at the time of their pericentric passages, implying that they did not evolve in the same environment as M33. These conclusions are generally upheld across all mass combinations explored. Varying the mass of M31 can affect whether Pisces is on first passage or if it has already completed a pericenter, as can be the case for M31 masses greater than the fiducial mass. Varying the mass of M33 does not change the conclusion that both dwarfs had a recent fly-by passage with M33, but changing the mass of M33 can affect how likely it is that these encounters occurred within the virial radius of M33.

It is still unknown whether the dwarfs were first accreted into the halo of M33 or M31, and whether the accretions occurred in tandem or as isolated events. The probability of two dwarfs having orbital poles aligned as closely as those of Pisces and And~XVI is $\leq 0.013$, further substantiating that these dwarfs are related rather than independently orbiting the M31-M33 system.

Future improvements in proper motion measurements for the dwarf galaxies and for M31, combined with more sophisticated orbital models, including improved M31 mass estimates, are needed to constrain further the accretion history of these dwarf galaxies and their relation to M31 and M33.

In rare cases, the dwarfs may have briefly transferred between the halos of M33 and M31. This phenomenon could complicate comparisons to surveys taking a satellite census around similar-mass host galaxies, and for example, reclassifying Pisces as an M33 satellite would have implications for interpreting quenched fractions and satellite statistics in surveys like ELVES and SAGA \citep{Carlsten_2022, Mao_2024}.

\appendix

\section{Figures Illustrating Orbits Across Different Combinations of M33 and M31 Masses}
\label{sec:appendix}
\begin{figure*}
    \centering
    \includegraphics[width=\linewidth]{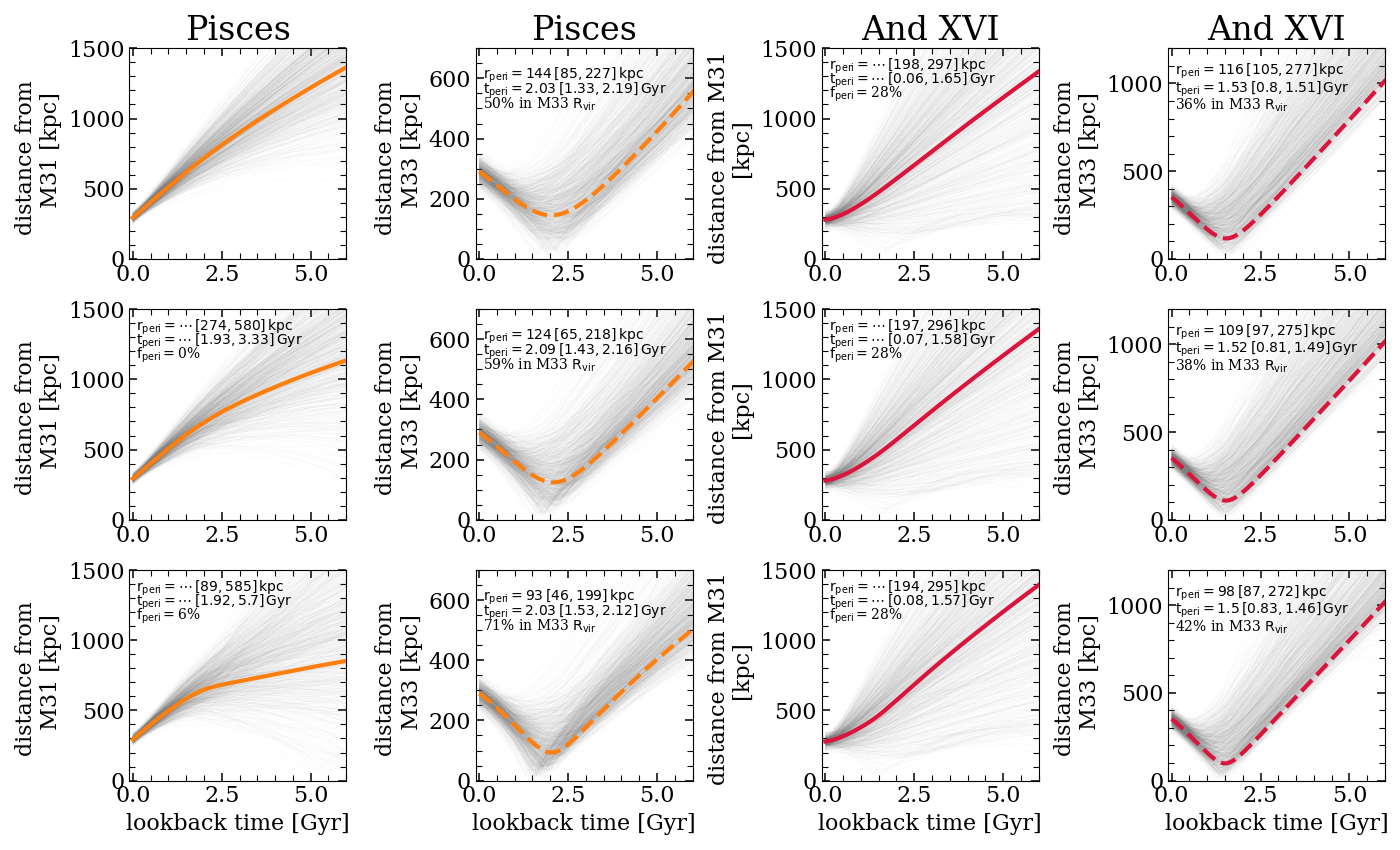}
    \caption{Orbits of Pisces and And~XVI for a fixed M31 mass of $1.5\times10^{12}\, M_{\odot}$. The first and third columns show orbits of each dwarf relative to M31, while the second and fourth columns show orbits relative to M33. Second and fourth columns also indicate the distance and time of pericenter, followed by the [15.9, 84.1] percentile, as well as, the fraction of orbits where the pericenter distance is within the virial radius of M33. From top to bottom, the rows correspond to varying M33 masses such that the top row represents an M33 mass of $1\times10^{11}\,M_{\odot}$, the middle row is for the fiducial M33 mass of $2.5\times10^{11}\,M_{\odot}$, and the bottom for is for an M33 mass of $5\times10^{11}\,M_{\odot}$. The gray lines correspond to 1,000 orbital uncertainties calculated for each combination of M33 and M31 mass.}
\label{fig:mass_combo1}
\end{figure*}

\begin{figure*}
    \centering
    \includegraphics[width=\linewidth]{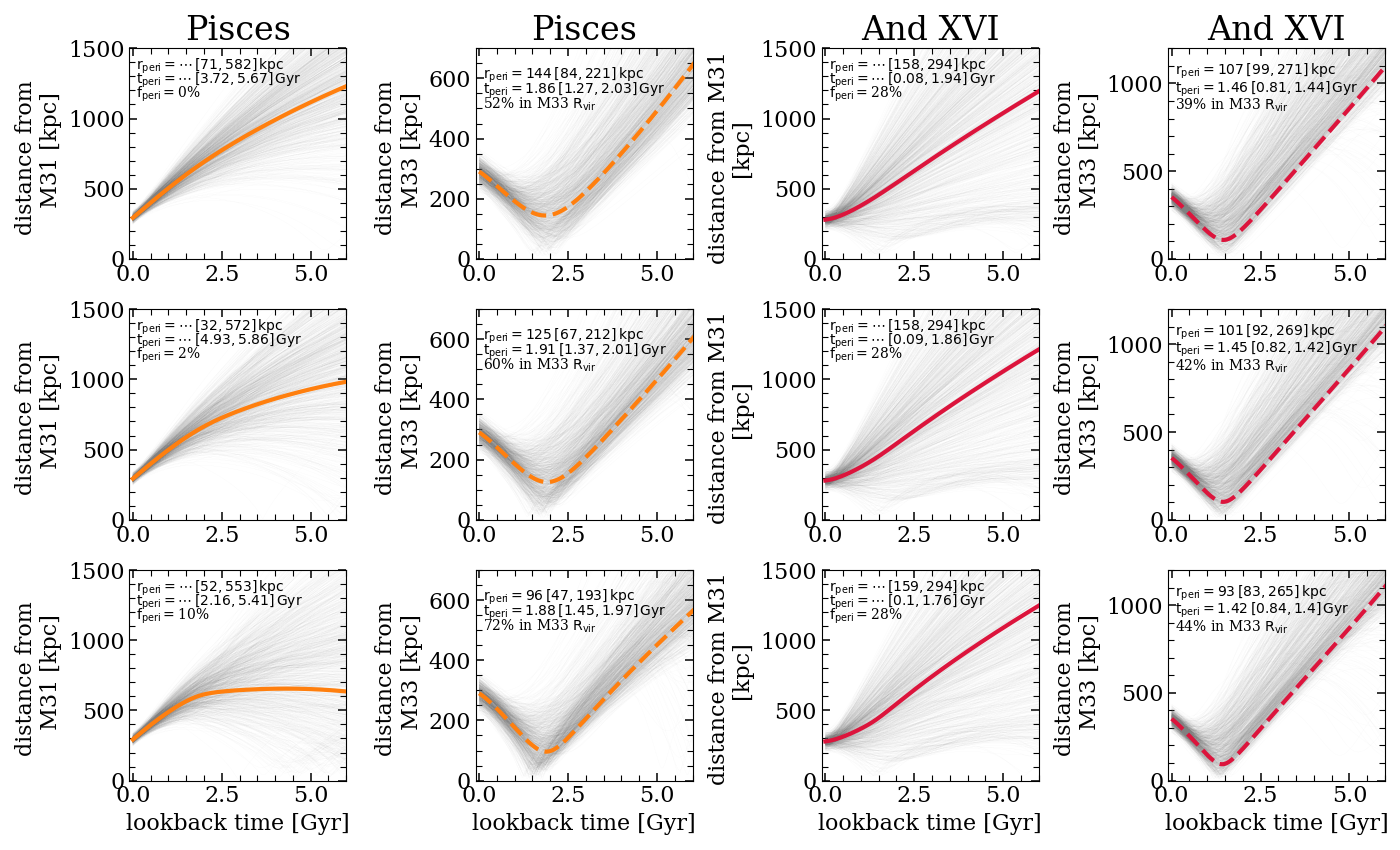}
    \caption{Same as Figure \ref{fig:mass_combo1} but assuming the fiducial M31 mass of $2\times10^{12}\, M_{\odot}$.}
\label{fig:mass_combo2}
\end{figure*}

\begin{figure*}
    \centering
    \includegraphics[width=\linewidth]{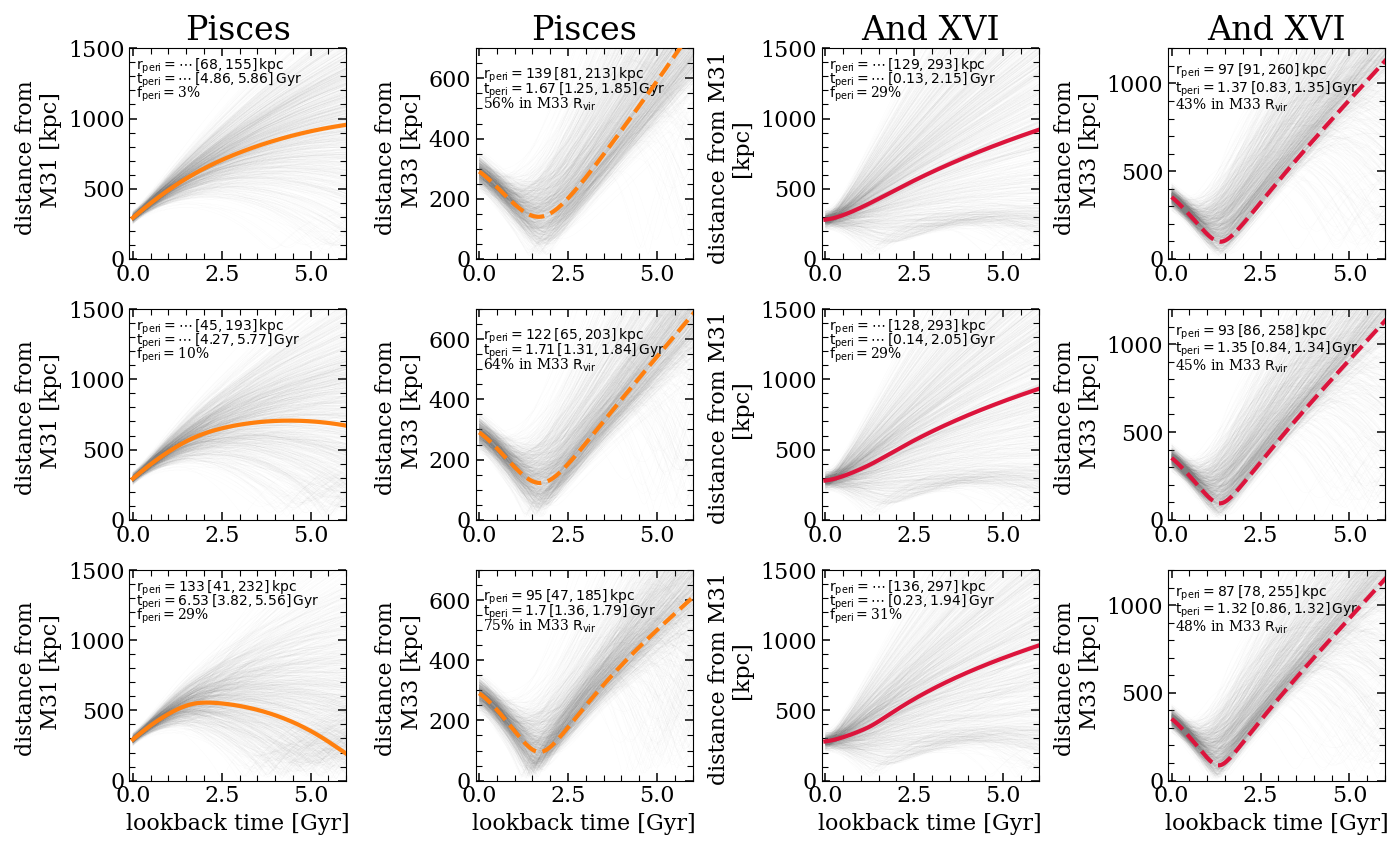}
    \caption{Same as Figure \ref{fig:mass_combo1} but assuming an M31 mass of $3\times10^{12}\, M_{\odot}$.}
\label{fig:mass_combo3}
\end{figure*}

\begin{acknowledgements}
NASA financially supported EP through the Hubble Fellowship grant \# HST-HF2-51540.001-A awarded by Space Telescope Science Institute (STScI). STScI is operated by the Association of Universities for Research in Astronomy, Incorporated, under NASA contract NAS5-26555. Additional support for this work was provided by NASA through grants for programs GO-15911, 16273, and 17174 from STScI. This work was based on the NASA/ESA HST observations and obtained from the Data Archive at STScI. 

The authors thank Alessandro Savino and Nico Garavito-Camargo for their helpful comments, which improved the manuscript.

\end{acknowledgements}

\software{Numpy \citep{numpy},
  SciPy \citep{SciPy-NMeth},
  Matplotlib \citep{matplotlib},
  IPython \citep{ipython},
  Jupyter \citep{jupyter}, 
  Astropy \citep{astropy:2013, astropy:2018, astropy:2022}, 
  Gala \citep{gala}}

\bibliography{refs}{}
\bibliographystyle{aasjournal}
\end{document}